PHELEX: Two Series of Measurements in the Search for Dark Photons.

A. Kopylov, I. Orekhov, V. Petukhov and A. Solomatin
Institute for Nuclear Research of RAS, 117312 Moscow, Prospect of 60 Anniversary of
October Revolution 7A, Russia

Abstract.

The results of measurements in two series, each series of 4 Runs and ≈ 50 days each Run, are presented. A multi-cathode counter with an iron cathode is used in these measurements. In both series we have observed some excess in count rates in sidereal time while no excess has been observed in terrestrial time. In the first series from 27.12.22 to 24.07.23 we have observed the excess in the period between 8-00 and 12-00 of sidereal time with a confidence level > 6 σ, in the second series – from 21.09.23 to 03.04.24 – in the period between 18-00 and 22-00 of sidereal time with a confidence level > 4 σ. The fact that in terrestrial time we do not observe a similar effect as in sidereal time suggests that probably this effect is of galactic origin and that this may be caused by dark photons. Future steps are suggested to validate these results.

**1. Introduction.**

The idea of this experiment and the results obtained in PHELEX (PHoton ELectron EXperiment) are presented in Ref. 1 – 13. We use a counter specially designed as an instrument to search for dark photons. It is a gaseous proportional counter with 3 cathodes [11], see Figure 1. The sensitivity of the method is proportional to the surface area of the outer metal cathode which ranges from 0.2 to 0.3 m$^2$. The typical diameter of the cathode is 140 to 200 mm and the length is approximately 500 mm. This method is mostly sensitive to masses of dark photons between 10 and 40 eV where the quantum efficiency of the metal cathode is high for the emission of single electron during conversion of dark photon at the cathode's surface. Near the Sun the density of dark matter is estimated to be 0.4 Gev/cm$^3$ [14]. It means that here we will have between 10 and 40 million particles of dark photons in 1 cm$^3$ if all dark matter is composed of dark photons with these masses. For comparison, we have only ≈ 150 relic neutrinos and ≈ 500 relic photons in 1 cm$^3$. Because they have mass, dark photons have a longitudinal mode and probably there is some interaction between dark photons. If so, they can be polarized at least through a longitudinal mode. Then the effect from this can be detected by our counter because it has a directionality. Dark photons with an E-field vector along the axis of the counter will not produce the effect. A maximal effect will be produced when the vector of the E-field is perpendicular to the axis of the counter [6]. Due to the rotation of the Earth, one should observe diurnal variations. The effect should depend on the orientation of the counter and on the geographical latitude of the site where it is located. For the Moscow site with a geographical latitude of 55º45' N, where our detector is located, the calculated diurnal variations are presented in Figure 2. Here 12-00 is taken at the moment when the vector of the E-field is in the plane of the Moscow meridian. If the vector of the E field is oriented by the angle relative axis of the Earth between 20º and 50º one should observe a maximum in the count rate during several hours depending upon the background count rate.

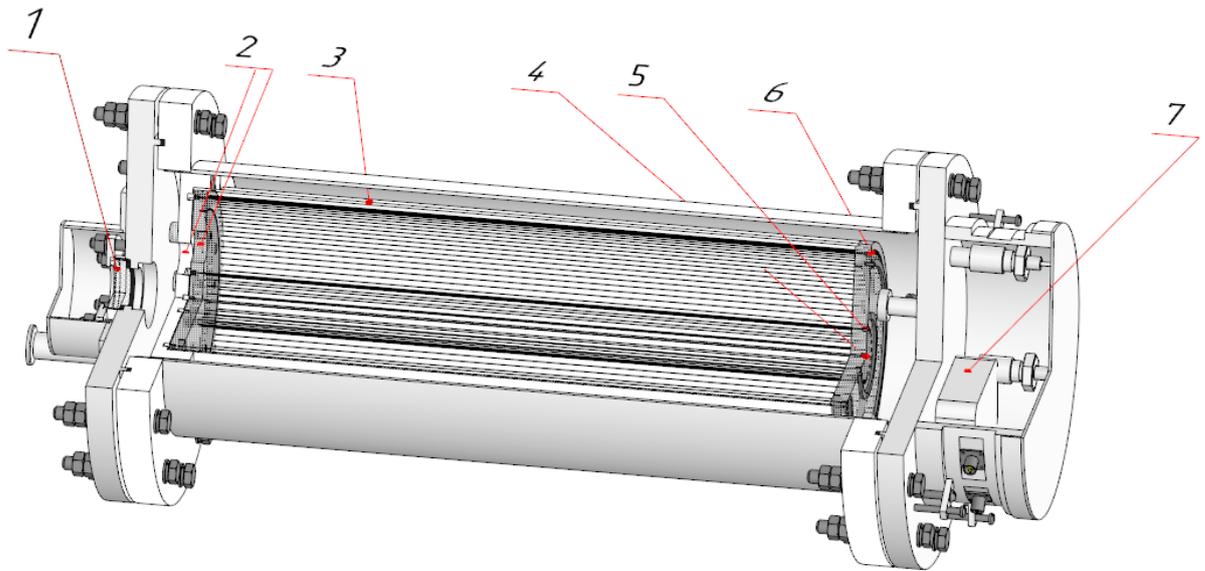

Figure 1. The multi-cathode counter. 1 – fused silica window for calibration, 2 – second and third windows for calibration, 3 – an iron cathode, 4 – anode, 5 – first cathode, 6 – second cathode, 7- preamplifier

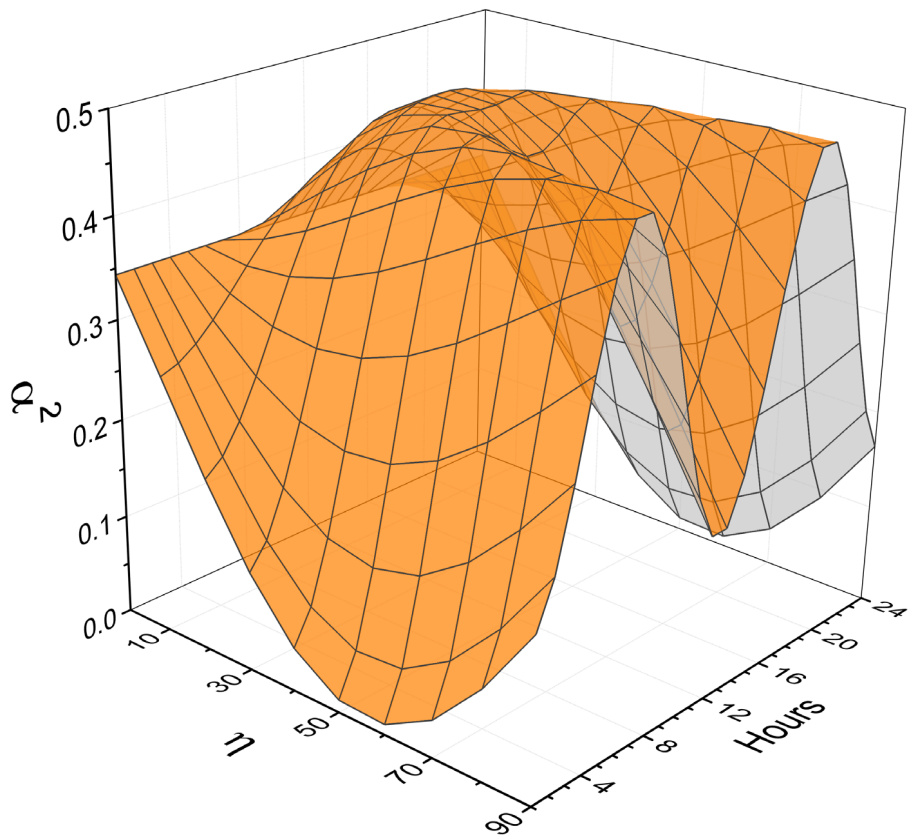

Figure 2. Diurnal variation calculated for counter placed in the laboratory at Moscow, Russia, in a horizontal position with an axis oriented along the meridian. On the vertical axis: $\alpha = \langle\cos^2\theta\rangle$, $\theta$ – the angle between the vector of E-field of dark photon and the surface of the cathode of the counter, $\eta$ – the angle between the vector of E-field of dark photon and the axis of Earth.

## 2. The results of measurements.

We have performed measurements by two series in the search for this effect. In both series we used a counter with an iron cathode with a diameter of 140 mm and a length of 500 mm. Figures 3 to 6 show the results obtained. As one can see in Fig. 3, we have observed a clear excess of count rates in the first series of measurements from 27.12.22 to 24.07.23 in the temporal interval between 8-00 and 12-00 of sidereal time in all four Runs. To assess the probability that this is the result of pure statistical fluctuation we have used an expression (1) for 4 -hours intervals (two points in each Run) as in [1]:

$$p = 12 \prod_{i=1}^{2n} \left(0{,}5\, erfc\left(\frac{x_i}{\sqrt{2}}\right)\right) \qquad (1)$$

Here: n is the number of Runs, and $x_i$ is the deviation of an i-th point in σ from the average value. For the interval between 8-00 and 12-00 we have found p = 7.6x10$^{-10}$ which corresponds to confidence level > 6 σ.

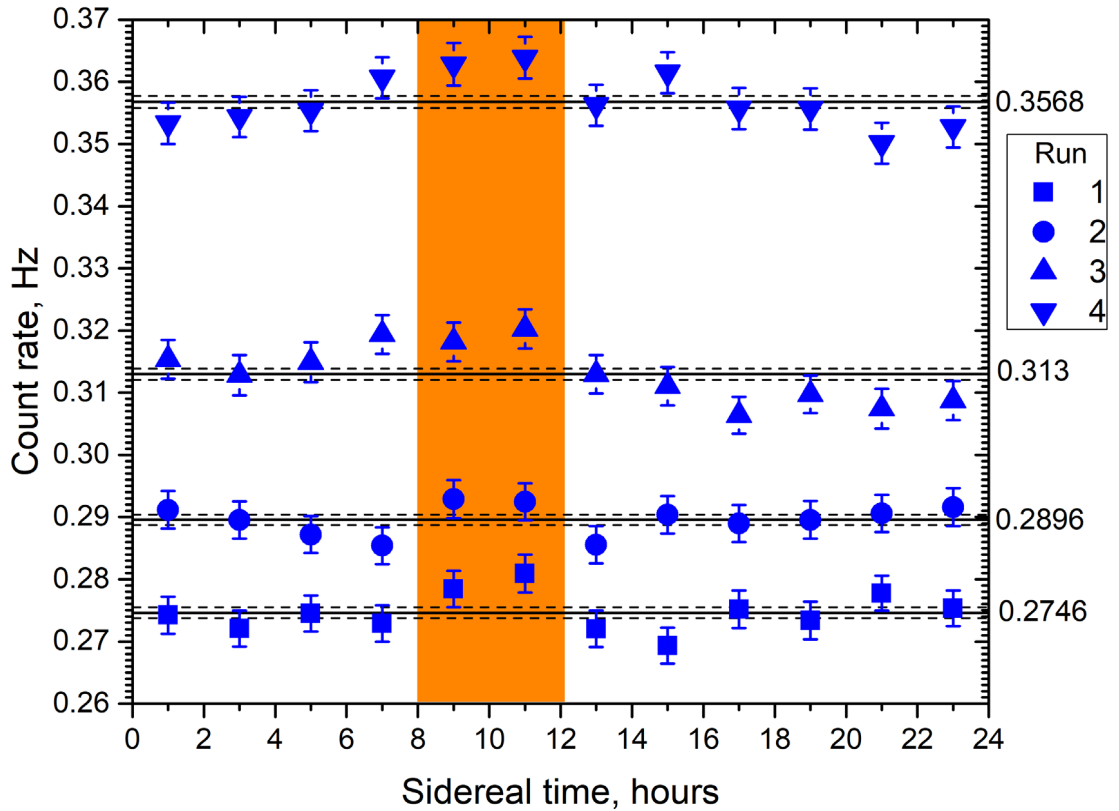

Figure 3. Diurnal variation in sidereal time in 4 Runs of the first series of measurements.

No similar effect (that one observes the excess at the same temporal period in all four Runs) has been observed in terrestrial time as one can see in Fig. 4. One can see some excess of counts in green zones of the third and fourth Runs what is expected because the events which contribute to the interval between 8-00 and 12-00 in sidereal time should belong to these green zones in terrestrial time. These green zones have a duration of not 4 h but ≈ 7 h 20 min because of the time shift between sidereal and terrestrial times during ≈ 50 days of measurement [1]. So, the effect observed within a 4-hour interval in Fig. 3 should be distributed within a broader interval of ≈ 7h 20 min in Fig. 4. One can see also from Fig. 3 that here the effect is more pronounced in the third and fourth Runs than in the first and second ones. This may explain why we observe in Fig 4 some excess in the third and fourth Runs; but do not see it in the first and second Runs.

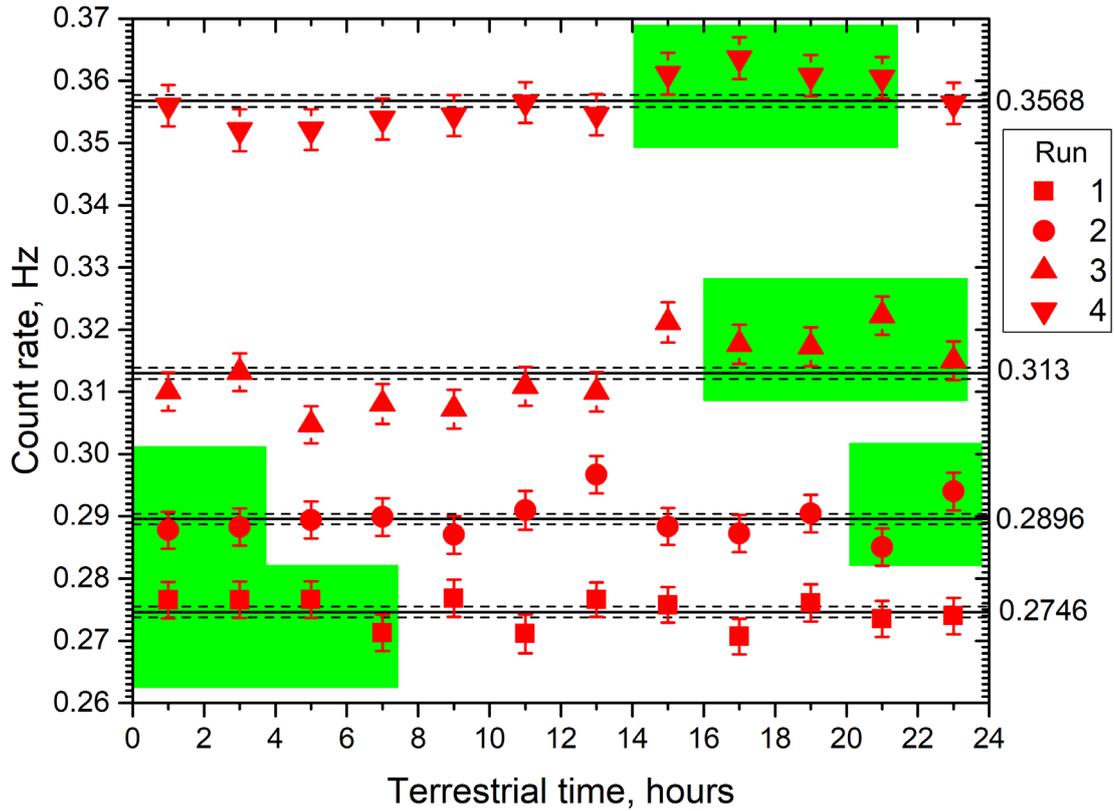

Figure 4. Diurnal variations in terrestrial time in 4 Runs of the first series of measurements. Green zones denote temporal intervals in terrestrial time that correspond to the interval between 8-00 and 12-00 of sidereal time in Fig. 3.

In the second series of measurements from 21.09.23 to 03.04.24 we also have observed a similar effect. As one can see from Fig. 5, we observe the excess in the interval between 18-00 and 22-00 of sidereal time in all four Runs. The probability that this is the result of pure statistical fluctuation was found to be $9.1 \times 10^{-6}$ which corresponds to C.L. > 4 σ. No similar effect has been seen in terrestrial time as one can see from Figure 6. To demonstrate how good is our criterium let's take an interval between 10-00 and 14-00 of sidereal time on Fig.5, which also contains scattered points and looks like a good candidate for possible effect. The probability for this interval is $3.0 \times 10^{-2}$ with C.L. < 2 σ. It is just noise. Let's take another interval from Fig. 6 between 0-00 and 04-00 which looks very perspective for the effect. Here we have the probability $7.0 \times 10^{-4}$ with C.L. < 3.5 σ. Thus, we can conclude that our criterium is reliable and all events with C.L.< 3.5 σ can be classified as noise while events with C.L. > 4 σ can be classified as candidates for the effect.

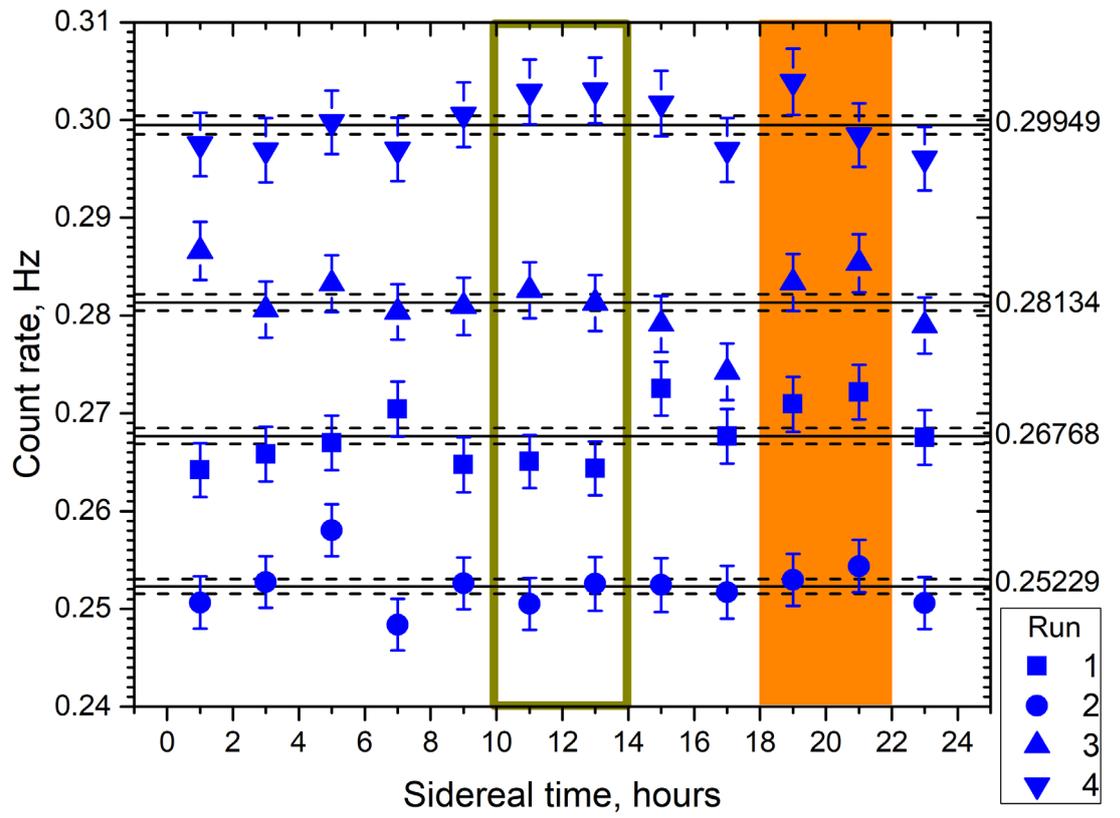

Figure 5. Diurnal variation in sidereal time in 4 Runs of the second series of measurements.

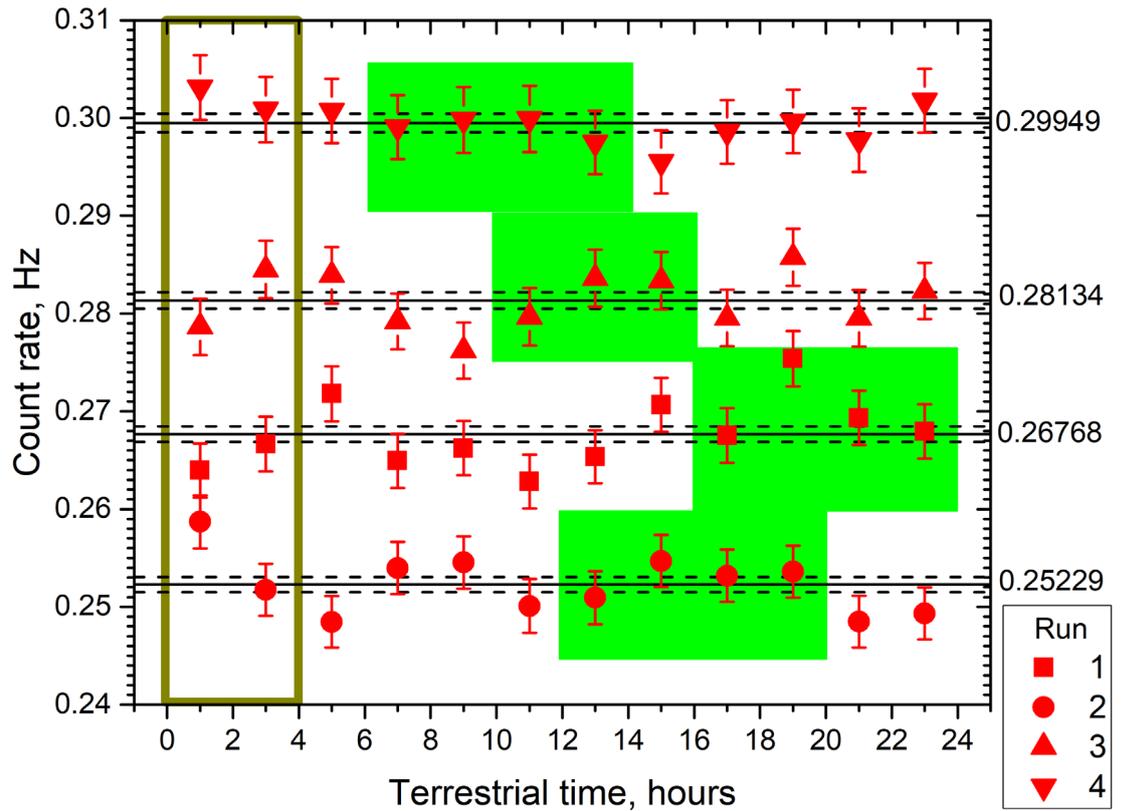

Figure 6. Diurnal variations in terrestrial time in 4 Runs of the first series of measurements. Green zones denote temporal intervals in terrestrial time that correspond to intervals between 18-00 and 22-00 of sidereal time in Fig. 5.

From here we can deduce the following. We have observed a clear excess of count rate in sidereal time in both series of measurements. In the first series it is in the temporal interval between 8-00 and 12-00 while in second series it is in the interval between 18-00 and 22-00. No similar excess in all 4 Runs in terrestrial time has been observed. The fact that temporal intervals are different tells that if we observe real effects, the vector of polarization from time to time changes its direction. The Sun is moving through space with a velocity of about 230 km/s. In 200 days, it flies $4 \times 10^9$ km. The interval between the first and second series was about two months. This is another billion km along the orbit around the center of the Milky Way. Along this way the vector of the E-field presumably has changed its direction. Why is the vector not changing its direction within 200 days? Probably it is changing. In the first series of measurements, as one can see in Fig.3, the effect is more pronounced in the third and fourth Runs than in the first and second ones. Probably this is the result of some change of the vector's direction. Certainly, all these inferences are of interest only if the effect was really observed. Although we have confidence levels > 6 σ in the first series and > 4 σ in the second series we can't be sure that we do observe the effect until we will see that similar effect is observed with several detectors working simultaneously. The effect should depend upon the orientation of the counter, on the geographical latitude and longitude of the site, it should not be observed in configuration 2, when the potential of the second cathode blocks electrons emitted from the metallic cathode to reach anode and possibly it also depends upon the material of the cathode etc. [2, 6]. For example, if we have second detector with the same orientation but located at the site near Tomsk in Siberia with geographical latitude close to the one of Moscow but with a 4-hour difference relative to Moscow, then we should observe the effect at Moscow the same as in Tomsk but with a 4-hour delay.

### 3. Conclusions.

In both series of measurements, we have observed in sidereal time some excess in count rate during 4 hours in all 4 Runs. In the first series it was observed with C.L. > 6 σ, and in the second series – with C.L. > 4 σ. The criterium used in our analyses is working well. Comparison with other temporal intervals in our analyses clearly indicates that for C.L. < 3.5 σ we can define the event as noise. For C.L. > 4 σ we can define the event as candidate for the effect. In terrestrial time we do not observe a similar effect as in sidereal time. This fact suggests that probably this effect is of galactic origin and that this may be caused by dark photons. To check this hypothesis, we are planning to use a second detector with the same orientation placed at the same site. If both detectors working simultaneously show the same picture this will provide an additional argument in support of this hypothesis.

**References.**

1. A.Kopylov, I.Orekhov, V.Petukhov, Latest Results on the Search of Dark Photons with a Multicathode Counter, Physics of Atomic Nuclei, 2023, Vol.86, No. 6, pp 1009-1013
DOI: 10.1134/S106377882306011X
2. A.Kopylov, I.Orekhov, V.Petukhov, PHELEX: Present Status, Moscow University Physics Bulletin, 2022, Vol. 77, pp, 315-318. DOI: 10.3103/S0027134922020539
3 . A.Kopylov, I.Orekhov, V.Petukhov, Diurnal Variations of the Count Rates from Dark Photons, Particles 2022, 5, 180-187. https://doi.org/10.3390/particles5020016
4. A.Kopylov, I.Orekhov, V.Petukhov, Multi-cathode counter as a detector of dark photons , Physics of Atomic Nuclei, 2022, Vol.85, No. 6, pp 1-9, (2022).
DOI: 10.31857/S0044002722060083
5. A.Kopylov, I.Orekhov, V.Petukhov, Present Status of the Experiment on the Search for Dark Photons by a Multi-Cathode Counter, Physics of Atomic Nuclei, 2021, Vol.84, No. 6, pp 860-865
6. A.Kopylov, I.Orekhov, V.Petukhov, On the possibility of observing diurnal variations in the count rate of dark photons using a multicathode counters, Physics of Particles and Nuclei, 2021, Vol.52, No.1, pp. 31 - 38
7. A.Kopylov, I.Orekhov, V.Petukhov, First results and future prospects with PHELEX, Journal of Physics, Conference Series, 1690 (2020) 012002, doi:10.1088/1742-6596/1690/1/012002
8. A.Kopylov, I.Orekhov, V.Petukhov, Results from a hidden photon dark matter search using a multi-cathode counter, JCAP, 07, 008 (2019)
9. A.Kopylov, I.Orekhov, V.Petukhov, Method of search for hidden photons of Cold Dark Matter using a multi-cathode counter, Physics of Atomic Nuclei, Vol.82, No. 9, pp 1-8, (2019)
10. A.Kopylov, I.Orekhov, V.Petukhov, Search for Hidden Photon Dark Matter using a Multi-Cathode Counter. Proceedings of the 4th International Conference on Particle Physics and Astrophysics (ICPPA-2018), Journal of Physics: Conference Series, 1390 (2019) 012066
11. A.Kopylov, I.Orekhov, V.Petukhov, A multi-cathode counter in a single-electron counting

mode, NIM A, 910, 164 (2018)

12. A.Kopylov, I.Orekhov, V.Petukhov, Tech. Phys. Lett 42, 102 (2016)

13. A.Kopylov, I.Orekhov, V.Petukhov, Adv. High Energy Phys., 2058372 (2016)

14. Marco Cirelli, Alessandro Strumia, Jure Zupan, Dark Matter, arXiv:2406.01705[hep-ph]